\documentclass[british]{article}
\usepackage{amsmath,amssymb,mathrsfs,bm,babel}

\title{Twisted Covariance and Weyl Quantisation}

\author{Gherardo Piacitelli\thanks{SISSA, Via Beirut 2--4, 34151, Trieste, Italia. Email: {\tt piacitel@sissa.it}}}
\newcommand{\bx}{\mathbf x}

\usepackage{hyperref}

\begin{document}

\maketitle

\begin{abstract}
In this letter we wish to clarify in which sense the tensor nature of
the commutation relations
\[
[\bx^\mu,\bx^\nu]=i\theta^{\mu\nu}
\]
underlying Minkowski spacetime quantisation
cannot be suppressed even in the twisted approach to Lorentz covariance. 
We then address the {\itshape vexata quaestio:} ``why~\(\theta\)?''

\end{abstract}

\section{Introduction}
We consider spacetime quantisation induced in a specific reference frame by commutation relations
of the form 
\[
[\bx^\mu,\bx^\nu]=i\theta^{\mu\nu}
\]
among the coordinates, for some fixed 
arbitrary choice of the real antisymmetric matrix \(\theta\). Ordinary
functions of the classical Minkowski spacetime are quantised according to a formally covariant
version of the Weyl prescription, as suggested in a more general (and fully covariant) context by
\cite{dfr}. According to a remark of \cite{ct}, the associated twisted product is form--invariant under a 
correspondingly twisted action of Poincar\'e covariance. This is usually interpreted as a 
fundamental breakdown of Lorentz covariance, embodied by the asserted non tensor character of
the invariant matrix \(\theta\).
 
In the next section we will instead unveil the hidden
tensor character of \(\theta\), which will necessarily emerge from
the interplay between twisted covariance and Weyl 
quantisation. The simple argument will rely on the assumption 
that all the observers adopt some {\itshape a priori}  unspecified 
Weyl quantisation as their quantisation
prescription. These results are described in more detail in \cite{piac}, and 
give a definitive answer to the conjecture raised in 
\cite{gb1} (see also \cite{gb}).

As a consequence of this remark, the twisted covariant approach 
can be recognised (see \cite{piac} for more details) as essentially equivalent to 
superposing a non invariant constrain on the fully covariant DFR model \cite{dfr}, 
determined by an arbitrary choice of the tensor \(\theta\) in an arbitrary frame of reference.
In the last section we comment on the implications of this choice.
 
Finally, in the appendix we briefly describe the formalism for dealing with 
symbols as functions of more general coordinates on \(\mathbb R^4\), while
mantaining the meaning of the quantisation prescription.

\section{Covariance}

Fix an observer, say Jim, in his own frame of reference, once and for all. 
Given (in that frame) the commutation relations
\[
[\bx^\mu,\bx^\nu]=i\theta^{\mu\nu},
\]
the Weyl quantisation
\[
W(f)=\int dk\,\check f(k)e^{i k_\mu\bx^\mu}
\]
(where \(\check f(k)=(2\pi)^{-4}\int dx\,f(x)e^{-ik_\mu x^\mu}\)) induces a symbolic calculus
\[
W(f)W(g)=W(f\star_\theta g)
\]
in terms of a twisted product which  may be written in the form
\[
f\star_\theta g=m_\theta(f\otimes g)=m(F_\theta f\otimes g),
\]
where \(m\) is the ordinary pointwise multiplication (=~restriction to the diagonal set \(x=y\)), and
\[
(F_\theta f\otimes g)(x,y)=
\frac{4}{|\det\theta|}\iint dadb\,f(x+a)g(y+b)
e^{2ia_\mu(\theta^{-1})^{\mu\nu}b_\nu}
\]
fulfils \(F_\theta^{-1}=F_{-\theta}\); a degenerate \(\theta\) 
would require some {\itshape proviso}; see \cite{piac}.
\(\mathcal A_\theta\) will denote the Weyl algebra of symbols.

Let \((\alpha(L)f)(x)=f(L^{-1}x)\) be the action of  the Poincar\'e group on 
\(\mathcal A_\theta\), where
\(L=(\varLambda,a)\). According to a remark
of \cite{ct}
, we may deform the ordinary coproduct in the sense of Drinfeld twists,
thus obtaining a twisted action of the Poincar\`e transformations
\[
\alpha^{(2)}_\theta(L)=F_\theta^{-1}(\alpha(L)\otimes\alpha(L))F_\theta
\]
on \(\mathcal A_\theta\otimes\mathcal A_\theta\),
which is compatible with the twisted product in the sense that
\begin{equation}\label{TC}
m_\theta(\alpha^{(2)}_\theta(L) f\otimes g)=\alpha(L)m_\theta(f\otimes g);
\end{equation}
the above compatibility condition is called twisted covariance.

This fact is commonly interpreted as a fundamental breakdown of ordinary Lorentz covariance,
responsible of which should be 
the asserted non tensor character of the matrix~\(\theta\).

Already at a formal level, this view is at least questionable: with
\begin{equation}\label{it_is_a_tensor!}
{\theta'}^{\mu\nu}={\varLambda^\mu}_{\rho}{\varLambda^\nu}_{\tau}\theta^{\rho\tau};
\end{equation}
then the straightforward commutation rule
\[
(\alpha(L)\otimes\alpha(L))F_\theta=F_{\theta'}(\alpha(L)\otimes\alpha(L))
\]
entails that
\[
m_\theta(\alpha^{(2)}_\theta(L) f\otimes g)\equiv(\alpha(L)f)\star_{\theta'}
(\alpha(L)g),
\]
where of course \(\star_{\theta'}\) is the product twisted with the transformed
matrix. The above identity is then identical to the more appealing 
\[
(f\star_\theta g)'\equiv f'\star_{\theta'}g',
\]
where primed functions are obtained by means of ordinary Poincar\'e action.

In other words, the formalism of twisted covariance is completely 
equivalent to the formalism
where the Poincar\'e action is untwisted, and the matrix \(\theta\) 
is treated as a tensor.

Since however the relativity principle is {\itshape a priori}
broken  by the choice of
a particular \(\theta\) in a particular frame, formal covariance is void of 
meaning and cannot be taken alone as a guidance.
In order to decide which of the two formalisms is more tailored to the
conceptual framework, we need to rely on the physical interpretation
of \(i\theta\) as the commutator of the coordinates, and ask ourselves which
commutation rules are observed in a different frame.

We then adopt the point of view of twisted covariance and assume that
\(\theta\) is invariant (not a tensor). 
Consider another observer~---~Jane~---~in
the reference frame
connected to Jim's by \(L\). Jane also is doing physics and she writes down
her own Weyl quantisation \(W'(f)=\int dk\check f(k)e^{i k\bx'}\) in terms of the 
coordinates \(\bx'\) in her frame. We make no {\itshape a priori} assumptions 
on the commutation relations for \(\bx'\).

Now we use twisted covariance \eqref{TC}: we must
have
\[
W'(m_\theta(\alpha^{(2)}_\theta(L)f\otimes g)=W'(\alpha(L)f)W'(\alpha(L)g).
\]
We compute  
%
%
\begin{align*}
W'(m^{(2)}_\theta&(\alpha^{(2)}_\theta(L)f\otimes g))=\\
=\iint& dhdk\; e^{i(k+h)_\mu({\bx'}-a)^\mu}
e^{-\frac{i}{2}(\varLambda^{-1}h)_\mu\theta^{\mu\nu}(\varLambda^{-1}k')_\nu}\\
&\check f(\varLambda^{-1}h)\check g(\varLambda^{-1}(k')),\\
W'(\alpha^{(1)}&(L)f)W'(\alpha^{(1)}(L)g)=\\
=\iint& dhdk\;e^{-i(h+k)_\mu a^\mu}e^{i h_\mu{{\bx'}^\mu}}e^{i k_\mu{\bx'}^\mu}
\check f(\varLambda^{-1}h)\check g(\varLambda^{-1}k),
\end{align*}
from which (and the arbitrarity of \(f,g\)) the Weyl relations for the \({\bx'}^\mu\)'s are immediately recovered:
\begin{equation}\label{newyeil}
e^{i h_\mu{\bx'}^\mu}e^{i k_\mu{\bx'}^\mu}=e^{-\frac{i}{2} h_\mu{\theta'}^{\mu\nu}k_\nu}
e^{i (h+k)_\mu{\bx'}^\nu},
\end{equation}
where \(\theta'\) is given by \eqref{it_is_a_tensor!}.

Equation \eqref{newyeil} is the Weyl form of the relations
\[
[{\bx'}^\mu,{\bx'}^\nu]=i \theta'^{\mu\nu}.
\]
The intrinsic tensor nature of \(i\theta\) as the commutator of the 
coordinates is then established in the framework of twisted covariance.
This speaks in favour of the formalism of covariant twisted products and
untwisted Poincar\'e actions, which is simpler to deal with than twisted
covariance. Indeed, there is no evident reason why 
unprivileged observers should prefer to 
use different matrices for the commutation 
relations of the coordinates and for the associated twisted product.

\section{Why \(\theta\)?}

Let us first clarify the issues at hand by means of an elementary example 
(the ``Newtonian example'', in what follows). Consider the 
Newton laws, together with Galilei covariance and the relativity principle.
A non covariant modification of the theory could be obtained by  
complementing the three Newton laws with a criterion for
{\itshape a priori} selecting a non invariant set of solutions. This of course should be
done by assigning a rejection rule in some specific reference frame 
(Jim's frame, to fix ideas). In this case the equations
would be formally covariant: a different observer (Jane, say) agreeing with 
Jim's choice 
would translate the constrain on the allowed solutions in her own coordinates. 
For example, if Jim discards the solutions \((x(t),y(t),z(t))\) 
such that \(z(0)<0\), and if Jane's frame is rotated by \(180^\circ\) w.r.t.\ 
Jim's,  then Jane would discard solutions with \(z'(0)>0\) in her coordinates \(x',y',z'\).
While such a choice would be perfectly acceptable if motivated by contingent 
reasons external to the general theory (e.g.\ interest in some specific problem), 
the promotion of such a selection criterion to
a new fundamental law of mechanics would be highly questionable, since it would severely break 
the relativity principle; non invariant constrains 
are very bad candidates to be general laws.  For example, it would allow to give
an absolute criterion for classifying the observers. The first question one should ask  
would be: ``What's wrong with the discarded solutions?'' 

If one agrees on treating properly the Weyl quantisation in all reference frames,
the situation now resembles our Newtonian example: 
the formalism is essentially 
covariant, but we may classify the observers according to which \(\theta'\)
they see. Within a fully covariant theory, instead, 
all transformed \(\varLambda\theta\varLambda^t\)
should be available at once together with \(\theta\) to each observer, and in
particular to Jim (precisely 
like all initial positions of motions should be available to all observers in the 
Newtonian example).
The latter is precisely the point of view adopted by \cite{dfr}. 

Indeed, by carefully rephrasing our simple remark, it is possible to show \cite{piac} 
that, if Weyl quantisation is treated properly and \(\theta\) fulfils
the DFR stability condition, the formalism of twisted covariance 
is equivalent to the fully covariant  DFR formalism \cite{dfr}  up to discarding a huge,
non invariant set
of admissible localisation states: only localisation states \(\omega\) 
which are pure on the centre of the DFR algebra and 
such that \(\omega(-i[\bx^\mu,\bx^\nu])=\theta^{\mu\nu}\) in Jim's frame
are allowed for. 

We are thus facing a precise analogue of our non covariant 
modification of the Newtonian example.
Hence we ask the natural question:
``What's wrong with the discarded localisation states?''

\section*{Appendix}

In some applications, 
it may be useful to work with symbols of different coordinates. In this
appendix we develop the formalism accordingly.
 
Let \(x\) be the canonical coordinates of \(\mathbb R^4\) and 
\(\xi\)
other global coordinates\footnote{Namely \(A\ni\xi\mapsto x\in\mathbb R^4\)
is a surjective diffeomorphism, for some open domain \(A\subset\mathbb R^4\).} 
with domain \(A\). 
We wish to describe the twisted product of symbols as
functions of \(\xi\), instead of \(x\). 
We adhere to the standard abuse of notations according to which \(x,\xi\) 
are points and \(x(\cdot),\xi(\cdot)\) are the coordinate
maps.

The quantisation prescription for \(f=f(\xi)\) then becomes 
\[
W_{\theta;\xi}(f)=\int dk\; f(\xi(\cdot))\check{\phantom{|}}(k)e^{ik\bx},
\] 
where
\begin{gather*}
f(\xi(\cdot))\check{\phantom{|}}(k)=
\frac{1}{(2\pi)^4}\int dx\;f(\xi(x))e^{-ikx}=
\frac{1}{(2\pi)^4}\int_A \frac{d\xi}{j(\xi)}f(\xi)e^{-ikx(\xi)},\\
J(\xi)=\partial \xi/\partial x, \quad j(\xi)=|\det J(\xi)|.
\end{gather*}
Standard computations yield
\begin{eqnarray*}
\lefteqn{f_1(\xi(\cdot))\star f_2(\xi(\cdot))(x)=}\\
&=&\frac{4}{|\det\theta|}\iint\limits_{A\times A} \frac{d\xi_1d\xi_2}%
{j(\xi_1)j(\xi_2)}f_1(\xi_1)f_2(\xi_2)e^{2i(x-x(\xi_2))\theta^{-1}
(x(\xi_2)-x(\xi_1))},
\end{eqnarray*}
which coincides with the usual twisted product when \(\xi(x)\equiv x\).

We may then define a twisted product \(\star_{\theta;\xi}\) on the functions
of \(A\) by setting
\[
(f_1\star_{\theta;\xi}f_2)(\xi)=
f_1(\xi(\cdot))\star f_2(\xi(\cdot))(x(\xi)).
\]
By construction,
\[
W_{\theta;\xi}(f_1)W_{\theta;\xi}(f_2)=W_{\theta;\xi}
(f_1\star_{\theta;\xi}f_2)
\]

If 
\(f_1,f_2\) and \(\xi(\cdot)\) are analytic, then the 
Moyal expansion is available: 
\[
f_1(\xi(\cdot))\star f_2(\xi(\cdot))(x)=
\left. e^{\frac i2\frac{\partial}{\partial x^\mu}
\theta^{\mu\nu}\frac{\partial}{\partial y^\nu}}
f_1(\xi(x))f_2(\xi(y))\right|_{y=x},
\]
from which we deduce
\begin{gather}
(f_1\star_{\theta;\xi}f_2)(\xi)=
\left. e^{\frac i2\Theta(\xi)^{\mu\nu}\frac{\partial}{\partial \xi^\mu}
\frac{\partial}{\partial \eta^\nu}}
f_1(\xi)f_2(\eta)\right|_{\eta=\xi},\\
\Theta(\xi)^{\mu\nu}={J(\xi)^\mu}_{\mu'}{J(\xi)^\nu}_{\nu'}\theta^{\mu'\nu'}.
\end{gather}

The algebra of symbols so obtained is isomorphic to the algebra of canonical 
symbols.


\begin{thebibliography}{9}
\bibitem{dfr} S.\ Doplicher et al,
Phys.\ Lett.\ B {\bfseries 331} 39--44 (1994);
Commun.\ Math.\ Phys.\ {\bfseries 172},
187--220 (1995) \href{http://arxiv.org/abs/hep-th/0303037}{[arxiv:hep-th/0303037]}.
\bibitem{ct} M.\ Chaichian et al, 
Phys.\ Lett. {\bfseries B604}, 98--102 (2004) \href{http://arxiv.org/abs/hep-th/0408069}{[arxiv:hep-th/0408069]};
Phys.Rev.Lett. {\bfseries 94} 151602 (2005) [hep-th/0409096].
 Julius Wess, 
        unpublished lecture, Vrnja{\v c}ka Banja, August 29--September 2, 2003 
        \href{http://arxiv.org/abs/hep-th/0408080v2}{[arxiv:hep-th/0408080v2]}.
\bibitem{piac}G. Piacitelli, 
\href{http://arxiv.org/abs/0902.0575}{[arxiv:0902.0575]}.
\bibitem{gb1} J.\ M.\ Gracia--Bond\'ia, 
        Monografias de la Real Academia de Ciencias de Zaragoza {\bfseries 29},         129--140 
        \href{http://arxiv.org/abs/hep-th/0606107}{[arXiv:hep-th/0606107]}.
\bibitem{gb} J.\ M.\ Gracia--Bond\'ia, F.\ Lizzi, F.\ Ruiz Ruiz and P.\ Vitale,
        Phys.\ Rev.\ D {\bfseries 74} 025014 (2006)
        \href{http://arxiv.org/abs/hep-th/0604206}{[arXiv:hep-th/0604206]}.
\end{thebibliography}
\end{document}